%% file: kuehneretal2025.tex
\documentclass[draft]{agujournal2019}
\input{newcommands}
\usepackage{url} 
\usepackage{lineno}
\usepackage[inline]{trackchanges} 
\usepackage{soul}

\draftfalse

\journalname{JGR: Atmospheres}

\begin{document}

%
%


\title{Impact of
Non-Classical Gravity-Wave Dynamics
on
Middle-Atmosphere Mean Flow
and
Solar Tides}

%
%




\authors{T. Kühner\affil{1}, G. S. Völker\affil{1,2}, U. Achatz\affil{1}}

\affiliation{1}{Institut für Atmosphäre und Umwelt, Goethe Universität Frankfurt, Frankfurt am Main, Germany,}
\affiliation{2}{Leibniz Institute for Baltic Sea Research Warnemünde, Rostock, Germany}





\correspondingauthor{Ulrich Achatz}{achatz@iau.uni-frankfurt.de}



\begin{keypoints}
\item 
    An advanced gravity-wave parameterization has been used in a global model to investigate the effect of non-classical gravity-wave dynamics.
\item 
    Significant effects are found by generalizing wave fluxes, incorporating wave transience, and allowing for oblique gravity-wave propagation.
\item This affects both mean flows and solar tides, arguing for generalized gravity-wave parameterizations.
\end{keypoints}

%
%

%
%


\begin{abstract}
Conventional gravity-wave (GW) parameterizations neglect three aspects of GW dynamics. Instead of momentum and entropy fluxes they use Eliassen-Palm fluxes, thereby neglecting the possibility that resolved flow are not in geostrophic and hydrostatic balance. They neglect the transience of the GW field and of the resolved flow, by determining at every time step equilibrium profiles of GW fluxes that would result if the vertical GW propagation were instantaneous. Moreover, they also do not take into account lateral GW propagation and horizontal GW fluxes. Because the prognostic GW model MS-GWaM does not need to make these assumptions, it has been used in the global weather and climate code ICON to investigate their consequences for the simulation of monthly mean zonal mean flows and of solar tides. All three aspects are found to influence the simulation results significantly. Among those, transience and lateral propagation have the strongest impact. The mean circulation in the mesosphere and lower thermosphere is affected at all latitudes and in the stratosphere at high altitudes as well. This together with the directly modified GW forcing leads also to significant differences in the migrating and nonmigrating components of solar tides. Comparisons with tides retrieved from satellite data are most favorable if both aspects are taken into account. This argues for a correspondingly generalized treatment of GW dynamics in their parameterization, as an efficient alternative to GW permitting simulations.
\end{abstract}

\section*{Plain Language Summary}
Gravity waves (GW) are propagating oscillations of wind and temperature that take a significant influence on the large-scale circulation of air masses in the middle atmosphere, between about 15 and 100 km altitude. They also interact with and modify solar tides, planetary-scale waves that are forced by the daily cycle of the heating of the atmosphere by the sun. In the upper part of the middle atmosphere solar tides describe daily oscillations of wind by several 10 m/s, so that any observation there is strongly affected by them. Gravity waves are too short in scale for the grids of typical climate models to be able to represent them. Hence their effect must be accounted for by modules that represent their effect faithfully, so-called parameterizations. To the present day these parameterizations make a few assumptions that are not necessary anymore with a new generation of gravity-wave modeling. The prognostic GW parameterization has been used to assess the effect of dropping the conventional assumptions on the mean circulation and solar tides in the middle atmosphere. It is shown that the resulting modifications in the simulated flow are  significant. This argues for moving in GW parameterizations to the more general approach of MS-GWaM.

\section{Introduction}
Solar tides (STs) are planetary-scale atmospheric waves that are forced by the diurnal cycle of atmospheric heating and that are shaped by their propagation through the mean-atmosphere wind and temperature field. Energy conservation in an atmosphere with upwardly decreasing density entails especially large tidal amplitudes in the mesosphere and lower thermosphere (MLT). The responsible heating is primarily due to the absorption of infrared radiation by tropospheric water vapor and of solar ultraviolet radiation by stratospheric and lower mesospheric ozone, and to latent heat release in the troposphere \cite<e.g.>[]{chapman1970quantitative,forbes_atmospheric_1982,forbesAtmosphericTide21982,hagan_global-scale_1997,haganConnectionsDeepTropical2007,achatz2008mechanisms}.
In the spectral decomposition of the tidal signal with respect to longitude, the components traveling in a manner synchronous with the sun are termed migrating tides, while the rest constitutes the so-called nonmigrating tides. Overviews of STs in satellite data are given, e.g., by \citeA{panchevaAtmosphericTidesPlanetary2011,sridharan2019seasonal}. Nonmigrating tides are due in parts to nonmigrating heating components, but also to the modulation of migrating tides by planetary waves \cite{haganModellingDiurnalTidal2001,griegerDependenceNonmigratingDiurnal2004,oberheide2006diurnal,achatz2008mechanisms}.

Presumably as important is the interaction of STs with internal gravity waves (GWs). It modulates the GW field in the MLT, and it also influences the tidal signal
\cite<e.g.,>[]{mclandress1997sensitivity,mclandress2002seasonal, meyer1999gravity, ortland2006gravity, senfImpactMiddleatmosphereThermal2011b}. To the present day, however, this interaction is not understood well because climate models cannot resolve the full GW spectrum but rather must rely on GW parameterizations (GWPs). The limitations of GW parameterizations translate directly into uncertainties in the understanding of GW-ST interactions. \citeA{hinesInternalAtmosphericGravity1960} has suggested long ago that GWs are essential for the MLT. Known deficiencies in seasonal and climatological-mean wind and temperature fields in numerical weather forecasting and climate simulations have led to the development and implementation of the first GWP two decades later by \citeA{lindzen1981turbulence,palmer_alleviation_1986,mcfarlane_effect_1987}. Subsequent developments have been reviewed by \citeA{fritts2003gravity,kim2003overview,alexanderRecentDevelopmentsGravitywave2010}, and the present situation is discussed by \citeA{achatz2023multi,achatz_etal_2024}. Next to uncertainties with regard to GW sources and GW dissipation, open questions relate to three simplifications that conventional GWP use, often for reasons of efficiency: (i) In the so-called single-column approximation (1D), oblique GW propagation and horizontal GW fluxes are ignored. (ii) The steady-state assumption (SS) neglects all transience of large-scale flow and GW field, but rather assumes that GWs propagate from their source to the model top instantaneously, as if their group velocity were infinitely large. (iii) In the pseudomomentum approach (PM), the formulation of the GW impact and thermal impact is neglected, with the momentum fluxes being replaced by pseudo-momentum (Eliassen-Palm) fluxes. This would be appropriate if the resolved flow were in geostrophic and hydrostatic balance. The question arises how much these assumptions affect tides simulated by a climate model if applied in the GWP in use. Because STs have periods of at most a day, and because they are not balanced, it is at least not clear whether SS and PM can be applied, but also 1D might be a limitation.

Using prescribed fields of zonal-mean atmosphere and STs, \citeA{senfImpactMiddleatmosphereThermal2011b} have found that single-column and steady state modify the GW fluxes, the resulting GW drag, and the tidal forcing significantly. \citeA{ribstein2015interaction} and  \citeA{ribstein2016interaction} have extended this by allowing for the GW impact on a linear global model simulating the STs. They also found a strong effect of the steady-state assumption. Unfortunately, however, the metric terms in their prognostic equations for GW propagation have not been correct, so results should be taken with caution. This deficiency does not apply to the Multi-Scale Gravity-Wave Model (MS-GWaM) that has been gradually developed \cite{muraschko2015application,boloni2016interaction,wei2019efficient,jochum2025impact} and that has been implemented by \citeA{boloni2021toward,kim2021toward,kim_etal_2024, MSGWaMAThreeDimensionalTransientGravityWaveParametrizationforAtmosphericModels} into the German community weather and climate code ICON \cite{zanglICONICOsahedralNonhydrostatic2014}. It is a prognostic GW model, where the single-column, steady-state and pseudomomentum approach can be switched on and off. Because ICON resolves STs the coupled ICON/MS-GWaM system is a suitable testbed for the questions outlined above.

A corresponding study is reported here. For this purpose, the theoretical framework for the different MS-GWaM modes is presented in section \ref{sec: Theory MS-GWaM}, and the nomenclature for ST is defined in section \ref{sec: Solar Tides}. The results are shown in section \ref{sec: Results}, followed by a summary and final conclusions in section \ref{sec: Summary and Conclusions}.

\section{MS-GWaM}
\label{sec: Theory MS-GWaM}

The underlying multiscale theory of MS-GWaM has been developed by \citeA{bretherton1966propagation,grimshaw1975nonlinear,achatz2010gravity,achatz2017interaction}. A didactic introduction is given by \citeA{achatzAtmosphericDynamics2022}, and an overview, including the conventional simplifications SS, 1D, and PM, and the basic Lagrangian numerical approach, is given by \citeA{achatz2023multi}. While the above has been formulated for dynamics on an $f$-plane, \citeA{hasha_gravity_2008} serves the generalization of the eikonal equations for GW propagation on the sphere, and \citeA{MSGWaMAThreeDimensionalTransientGravityWaveParametrizationforAtmosphericModels} have implemented it into ICON/MS-GWaM. The following summarizes the most important aspects, beginning with the general formulation (3D) and then consecutively introducing the simplifications that are in line with conventional GWP.


\subsection{MS-GWaM 3D}
\label{subsec: Theory MS-GWaM 3D}

\subsubsection{Mean-Flow Impact on Gravity Waves}

Using spherical coordinates, i.e., longitude $\lambda$, latitude $\phi$ and radial distance $r$ from the center of the Earth, we first consider a locally monochromatic GW field with wavenumber $\mathbf{k}(\mathbf{r},t) = \mathbf{k}(\lambda,\phi,r,t) = k\mathbf{e}_{\lambda}+l\mathbf{e}_{\phi}+m\mathbf{e}_r$, using unit vectors $(\mathbf{e}_{\lambda},\mathbf{e}_{\phi},\mathbf{e}_r)$ pointing in the three coordinate directions. The GW local frequency is given by the dispersion relation $\omega\left(\mathbf{r},t\right)
     = \Omega\left[\mathbf{r},\mathbf{k}\left(\mathbf{r},t\right),t\right]$ with,
\begin{linenomath*}
\begin{equation}
     \Omega\left(\mathbf{r},\mathbf{k},t\right)
     =  \mathbf{k} \cdot \mathbf{U} (\mathbf{r},t)
        \pm
        \sqrt{
        \frac{N^2 (\mathbf{r},t)\left(k^2+l^2\right)+f^2 (\phi)\left[m^2+\Gamma^2(\mathbf{r},t)\right]}{\mathbf{k}\cdot \mathbf{k}+\Gamma^2(\mathbf{r},t)}
        }\ , \label{eq: dispersion relation 3D}
\end{equation}
\end{linenomath*}
where $\mathbf{U}\left(\mathbf{r},t\right)=U\mathbf{e}_{\lambda}+V\mathbf{e}_{\phi}$ is the mean-flow horizontal wind, $f\left(\phi\right)$ the Coriolis parameter, $N^2\left(\mathbf{r},t\right)$ local stratification, and $\Gamma\left(\mathbf{r},t\right)$ the scale-height correction. The eikonal equations predict the development of the local wavenumber along rays, defined by the local group velocity $\mathbf{c}_g = \nabla_{k} \Omega$ with components $(c_{g,\lambda},c_{g,\phi},c_{g,r}) = (\partial_k\Omega,\partial_l\Omega,\partial_m\Omega)$, i.e.,
\begin{linenomath*}
\begin{eqnarray}
    \dot{\lambda} = \left(\partial_t + \mathbf{c}_g \cdot \nabla\right)\lambda &=& \frac{c_{g,\lambda}}{r\cos\phi} \ , \label{eq: eikonal equations 3D eq3} \\
    \dot{\phi} = \left(\partial_t + \mathbf{c}_g \cdot \nabla\right)\phi &=& \frac{c_{g,\phi}}{r} \ , \label{eq: eikonal equations 3D eq4} \\
    \dot{r} = \left(\partial_t + \mathbf{c}_g \cdot \nabla\right)r &=& c_{g,r} \ , \label{eq: eikonal equations 3D eq5}
\end{eqnarray}
\end{linenomath*}
as,
\begin{linenomath*}
\begin{eqnarray}
    \dot{k} = \left(\partial_t + \mathbf{c}_g \cdot \nabla\right)k
    &=&-\frac{\partial_\lambda\Omega}{r\cos\phi}
    -\frac{k}{r}c_{g,r} 
    +\frac{k\tan \phi}{r}c_{g,\phi}\ , \label{eq: eikonal equations 3D eq0} \\
    \dot{l} = \left(\partial_t + \mathbf{c}_g \cdot \nabla\right)l
    &=& -\frac{\partial_\phi\Omega}{r}
    -\frac{l}{r}c_{g,r}-\frac{k\tan\phi}{r}c_{g, \lambda}\ , \label{eq: eikonal equations 3D eq1} \\
    \dot{m} = \left(\partial_t + \mathbf{c}_g \cdot \nabla\right)m
    &=& -\partial_r\Omega
    +\frac{k}{r}c_{g,\lambda}+\frac{l}{r}c_{g,\phi}\ . \label{eq: eikonal equations 3D eq2}
\end{eqnarray}
\end{linenomath*}
The development of the amplitude of the locally monochromatic field is predicted by the wave-action equation,
\begin{linenomath*}
\begin{equation}
    0 = \del_t \mathcal{A} + \nabla \cdot \left(\mathbf{c}_g \mathcal{A}\right) + \mathcal{S}\ ,
\end{equation}
\end{linenomath*}
where $\mathcal{A}$ is the wave-action density, and $\mathcal{S}$ is a source and sink term of the wave action. Numerical instabilities can occur due to caustics, i.e., areas where rays with different wavenumbers cross at a particular point in Euclidean space. Such situations arise from wave refraction, and they represent a breakdown of the locally monochromatic perspective, that is notably not in agreement with the typical atmospheric setting where one most often finds a full spectrum of GWs.

This problem can be bypassed with the spectral phase-space wave action density,
\begin{equation}
    \mathcal{N}=\mathcal{N}\left(\mathbf{r},\mathbf{k},t\right)=\sum_j \mathcal{A}_{j}\left(\mathbf{r},t\right)\delta\left[\mathbf{k}-\mathbf{k}_j\left(\mathbf{r},t\right)\right]\ , \label{eq: wad_spec}
\end{equation}
representing a superposition of different spectral components.  One can show that its development is predicted by,
\begin{eqnarray}
    0&=&\del_t \mathcal{N} + \nabla \cdot \left(\mathbf{c}_g \mathcal{N}\right) + \nabla_{\mathbf{k}}\cdot \left(\dot{\mathbf{k}}\mathcal{N}\right) + \tilde{\mathcal{S}} \nonumber \\
    &=& \left(\del_t + \mathbf{c}_g \cdot \nabla + \dot{\mathbf{k}} \cdot \nabla_{k} \right) \mathcal{N} + \tilde{\mathcal{S}}\ , \label{eq: phase-space wave-action conservation}
\end{eqnarray}
where $\tilde{\mathcal{S}}$ is the corresponding source and sink term. The nondivergence of the six-dimensional phase-space velocity,
\begin{equation}
    0 = \nabla \cdot \mathbf{c}_g + \nabla_k \cdot \dot{\mathbf{k}}\ , \label{eq: non-divergence phase-space}
\end{equation}
has been used. In the absence of sources and sinks, the spectral wave-action density is conserved along spectral rays propagating with the phase-space velocity.
A direct consequence of the nondivergence (\ref{eq: non-divergence phase-space}) is that phase-space volumes propagating along rays keep their 6-dimensional phase-space volume content.

\subsubsection{Gravity-Wave Impact on the Mean Flow: Direct and PM Approach}

\paragraph{Direct Approach}

The direct result from the multi-scale asymptotic theory is that the impact of GWs on the mean flow is described via GW momentum fluxes $\overline{\rho}\left<\mathbf{v}^{\prime} \mathbf{u}^{\prime}\right>$ and entropy fluxes $\left<\mathbf{u}^{\prime}\theta^{\prime}\right>$, so that,
\begin{eqnarray}
    \left.\del_t \mathbf{U}\right|_{\mathrm{GW}} &=& -\frac{1}{\overline{\rho}}\nabla \cdot \overline{\rho}\left<\mathbf{v}^{\prime} \mathbf{u}^{\prime}\right> + \frac{f}{\overline{\theta}}\mathbf{e}_r \times \left<\mathbf{u}^{\prime}\theta^{\prime}\right> , \label{eq: direct_momentum}\\
    \left. \del_t \theta\right|_{\mathrm{GW}} &=& -\nabla \cdot \left<\mathbf{u}^{\prime}\theta^{\prime}\right>\ .  \label{eq: direct_entropy}
\end{eqnarray}
Here, $g$ is Earth's gravity, and $\bar{\rho}$ and $\bar{\theta}$ are the reference-atmosphere density and potential temperature, respectively.
The bracket $\left< \cdot\right>$ is the phase average of the perturbed waves. The fluxes can be obtained from the spectral wave-action density by the integrals
\begin{eqnarray}
    \overline{\rho}\left<\mathbf{v}^{\prime} \mathbf{u}^{\prime}\right>
    &=& \int d^3k \frac{\hat{\omega}^2 \hat{\mathbf{c}}_g \mathbf{k}_h \mathcal{N} + f^2\left(\mathbf{e}_r \times \mathbf{c}_g\right)\left(\mathbf{e}_r \times \mathbf{k}_h \mathcal{N}\right)} {\hat{\omega}^2-f^2}\ , \label{eq: momentum flux}\\
    \left<\mathbf{u}^{\prime}\theta^{\prime}\right> &=& \int d^3k \frac{\overline{\theta}}{g \overline{\rho}}\hat{c}_{g,r}\frac{fN^2}{\hat{\omega}^2-f^2}\mathbf{e}_r \times \mathbf{k}_h \mathcal{N}\ , \label{eq: entropy flux}
\end{eqnarray}
where $\mathbf{k}_h = k \mathbf{e}_{\lambda} + l \mathbf{e}_{\phi}$ is the horizontal wavevector, and $\hat{\omega} = \omega - \mathbf{k}_h\cdot\mathbf{U}$ and $\hat{\mathbf{c}}_g = \mathbf{c}_g - \mathbf{U}$ are the intrinsic frequency and the group velocity, respectively. 

\noindent \paragraph{Pseudomomentum Approach}
In the conventional approach to GWP, one uses, instead of (\ref{eq: direct_momentum}) and (\ref{eq: direct_entropy}),
\begin{eqnarray}
    \left. \del_t \mathbf{U}\right|_{\rm{GW}} &=& - \frac{1}{\bar{\rho}}\nabla \cdot \left(\bm{\mathcal{G}}\mathbf{e}_{\lambda}+\bm{\mathcal{H}}\mathbf{e}_{\phi}\right)\ , \label{eq: pm_momentum}\\
    \left.\del_t \theta\right|_{\rm{GW}} &=& 0\ , \label{eq: pm_entropy}
\end{eqnarray}
where,
\begin{eqnarray}
    \bm{\mathcal{G}} &=& \int d^3k \hat{\mathbf{c}}_g k \mathcal{N}\ , \\
    \bm{\mathcal{H}} &=& \int d^3k \hat{\mathbf{c}}_g l \mathcal{N}\ ,
\end{eqnarray}
are the fluxes of the zonal and meridional components, respectively, of the GW pseudomementum $\mathbf{p}_h=\int d^3k \mathbf{k}_h \mathcal{N}$. These fluxes are often also called GW Eliassen-Palm fluxes. The reasoning behind this approach is that, if the mean flow is in hydrostatic and geostrophic balance, both (\ref{eq: direct_momentum}) - (\ref{eq: direct_entropy}) and (\ref{eq: pm_momentum}) - (\ref{eq: pm_entropy}) lead to the same prognostic gravity wave forcing for the quasigeostrophic potential vorticity equation and stream function. Hence they also also predict the same balanced winds and entropy. However, if the mean flow is not balanced, then there is no guarantee that the pseudomomentum approach predicts the correct mean flow. \citeA{wei2019efficient} show, that the response of the mean flow to an upward propagating GW packet can be predicted well by the direct approach, but not by the pseudomomentum approach. Because STs are not balanced, one might expect that they as well are better predicted by the direct approach than by the pseudomomentum approach.

\subsection{MS-GWaM 1D}


In the single-column approximation, GWs are only allowed to propagate vertically, i.e., they move in columns with fixed latitude and longitude. Horizontal mean-flow gradients are neglected as well. The  eikonal equations are further simplified by taking the limit $r\rightarrow\infty$. They take the form,
\begin{eqnarray}
    \dot{\lambda} = \dot{\phi} &=& 0 \label{eq: 1D_eikonal_lp}\ ,\\
    \dot{r} &=& c_{g,r}\ , \label{eq: 1D_eikonal_r}\\
    \dot{k} = \dot{l} &=& 0\ , \label{eq: 1D_eikonal_kl}\\
    \dot{m} &=& -\partial_r\Omega\ . \label{eq: 1D_eikonal_r}
\end{eqnarray}
The conservation property (\ref{eq: 1D_eikonal_kl}) is satisfied by keeping, for a selected set of spectral components, $k$ and $l$ constant. In the subdomain of each of these spectral components, the wave-action equation simplifies, again neglecting horizontal GW propagation, to,
\begin{equation}
    0 = \del_t \mathcal{N} + \del_r\left(c_{g,r} \mathcal{N}\right) + \del_{m} \left(\dot{m}\mathcal{N}\right) + \tilde{\mathcal{S}}
    = \del_t \mathcal{N} + c_{g,r} \del_r \mathcal{N} + \dot{m} \del_{m} \mathcal{N} + \tilde{\mathcal{S}}\ ,
\end{equation}
and the GW impact on the mean flow reads in the direct approach,
\begin{eqnarray}
    \left.\del_t \mathbf{U}\right|_{\rm{GW}} &=& -\frac{1}{\bar{\rho}}\del_r\left(\bar{\rho}\left<w^{\prime}\mathbf{u}^{\prime}\right>\right)\ , \label{eq: impact of gw on mean flow single-column eq1} \\
    \left.\del_t \theta \right|_{\rm{GW}} &=& 0\ , \label{eq: impact of gw on mean flow single-column eq0}
\end{eqnarray}
while in the pseudomomentum approach, the mean-flow impact impact becomes
\begin{equation}
    \left.\del_t \mathbf{U}\right|_{\rm{GW}} = - \frac{1}{\bar{\rho}}\del_r \left(\mathcal{G}_r\mathbf{e}_{\lambda} + \mathcal{H}_r\mathbf{e}_{\phi}\right)\ ,.\label{eq: impact of gw on mean flow single-column pm}
\end{equation}
The work of \citeA{boloni2021toward} and \citeA{kim2021toward} is recommended. Both describe and evaluate MS-GWaM 1D with PM approach.

\subsection{MS-GWaM SS}

Together with single-column, the steady-state approach leads to the conventional setup of GWP. It neglects mean-flow transience, i.e., one assumes $\partial_t\Omega = 0$. As a consequence of this one has,
\begin{equation}
    \dot{\omega} = 0\ ,
\end{equation}
which is solved by assuming constant frequency $\omega$, so that at a given time $t$, given horizontal location $(\lambda,\phi)$, and for given horizontal wave numbers $(k,l)$, the frequency,
\begin{equation}
    \omega = \Omega\left[\lambda,\phi,r,k,l,m(\lambda,\phi,r),t\right]
\end{equation}
can be solved for the vertical wave number $m$. Steady state also neglects GW transience, i.e., one assumes $\partial_t\mathcal{N} = 0$, so that the wave-action equation takes the diagnostic form,
\begin{equation}
    0 = \del_r\left(c_{g,r} \mathcal{N}\right) + \del_{m} \left(\dot{m}\mathcal{N}\right) + \tilde{\mathcal{S}} \ .
\end{equation}
Because the wave-action density must vanish at $m\rightarrow\pm\infty$ this integrates to,
\begin{equation}
    0 = \del_r \int dm\, c_{g,r} \mathcal{N} + \int dm\, \tilde{\mathcal{S}} \ ,
\end{equation}
or, for a spectral component with given $(k,l)$,
\begin{equation}
    0 = \del_r \left( c_{g,r} \mathcal{A} \right) + \mathcal{S} \ .
\end{equation}
This is solved for the wave-action density $\mathcal{A}$ given some boundary value at a chosen launch-level altitude. Using wave-action density and wave numbers, one finally obtains the GW fluxes that determine the GW impact on the mean flow. In the PM framework, this entails that the mean flow can only be affected in the presence of non-zero sources and sinks. Transience and 3D break this non-acceleration property.

\section{Solar Tides: Diagnostics and Nomenclature}
\label{sec: Solar Tides}
In the definition used here, STs are diagnosed as the average diurnal cycle of all dynamical fields, with the mean subtracted. Here we will take the average of all days of a given month. Let for any variable $Y_{\mathrm{ST}} (\lambda,\phi,r,t)$ be the derived tidal signal. It has become common to Fourier decompose $Y_{\mathrm{ST}}$ in longitude and time as \cite<e.g.,>[]{forbes2006troposphere, ribstein2015interaction, ribstein2016interaction},
\begin{equation}
     Y_{\mathrm{ST}} (\lambda,\phi,r,t)
     =
     \sum_{n=1}^{\infty}\sum_{k\in \mathbb{Z}}
     \widetilde{Y}_{\mathrm{ST}}\left(n,k,\phi,r\right)
     \cos\left[n\Omega_{\mathrm{T}} t + k\lambda + \phi_{n,k}(\phi,r)\right]
     \ , \label{eq: Definition solar tides mathematical}
\end{equation}
where $\Omega_{\mathrm{T}} = 2\pi/24\mathrm{h}$ is Earth's rotation rate, and $\phi_{n,k}$ and $\widetilde{Y}_{\mathrm{ST}}\left(n,k\right)$ are phase and amplitude of the solar tide component (also termed tidal component or tidal signal) $(n,k)$. The index $n = 1,\ 2,\ 3, \ldots$, indicates the tidal component with periods of $24\ \mathrm{h},\ 12\ \mathrm{h},\ 8\ \mathrm{h}, \ldots$, i.e., the diurnal, semidiurnal, terdiurnal,$\ldots$, solar tide. The index $k$ indicates the horizontal wavenumber of the ST component. The corresponding (zonal) phase velocity is,
\begin{equation}
    c_{p} = -\frac{n\Omega_T}{k}\ ,
\end{equation}
such that \textit{eastward} propagation corresponds to a negative zonal wavenumber ($k<0$), while \textit{westward} propagation corresponds to a positive zonal wavenumber ($k>0$). Tides characterized by a vanishing horizontal wavenumber $k=0$ are referred to as \textit{stationary} or \textit{standing} tides \cite{chapman1970quantitative}. If $k=n$, the phase velocity reduces to $c_{p} = -\Omega_{\mathrm{T}}$, indicating a westward-propagating solar tide with a phase velocity in agreement with the frequency of the Earth's rotation rate. These tides are referred to as \textit{migrating} tides and are sun-synchronous, i.e., following the sun's motion. Tides with $k \neq n$ are referred to as \textit{nonmigrating} tides. Solar tides are defined as the sum of its migrating and nonmigrating ST components. For simplicity, all ST components posses a shortcut $\mathrm{NPK}$. The shortcut $N \in \left\{\mathrm{D},\ \mathrm{S},\ \mathrm{T}\right\}$ denotes the diurnal, semidiurnal, or terdiurnal component with periods of 24h, 12h, and 8h, respectively. $\mathrm{P} \in \left\{\mathrm{W},\ \mathrm{E}\right\}$ denotes westward or eastward propagation. The shortcut $\mathrm{K} \in \mathbb{N}$ is the absolute value of the zonal wavenumber. Standing ST with $K =0$ are denoted by $\mathrm{N0}$, since they propagate neither eastward nor westward.

\section{Results}
\label{sec: Results}
For the data to be analyzed, ensembles of ICON/MS-GWaM integrations have been obtained in the various MS-GWaM setups (SS, 1D, 3D, 3D with PM). Each ensemble member has been initialized on Nov 1st of its given year, and the integrations extend until Dec 31st. Data has been stored every 3h, and only the December data has been used for the analysis. Monthly means have been obtained as well as mean diurnal cycles (composites), from which the diurnal, semidiurnal, and terdiurnal ST components have been obtained via Fourier decomposition as outlined above.

Following \citeA{kim2015t}, we diagnose, in the comparison of two approaches, statistical significance at level 0.05,
\begin{align}
    t\cdot \sqrt{\frac{\mathrm{Var}\left(\overline{X+Y}\right)}{N}+\frac{\mathrm{Var}\left(\overline{X+Y}\right)}{N}} < \left|\overline{X}-\overline{Y}\right|\,  \label{eq: significane test different data points}
\end{align}
with two ensembles $X$ and $Y$ having $N$ and $M$ members, respectively, a t score which is $t=1.74588$ or $t=1.73961$, due to ensembles having $18$ or $19$ total members, respectively, and a weighted variance,
\begin{align}
    \mathrm{Var}\left(\overline{X+Y}\right) = \frac{\left(N-1\right)\mathrm{Var}\left(X\right)+\left(M-1\right)\mathrm{Var}\left(Y\right)}{N+M-2}\ .
\end{align}

\begin{figure}[h]
    \centering
    \includegraphics[width=\textwidth]{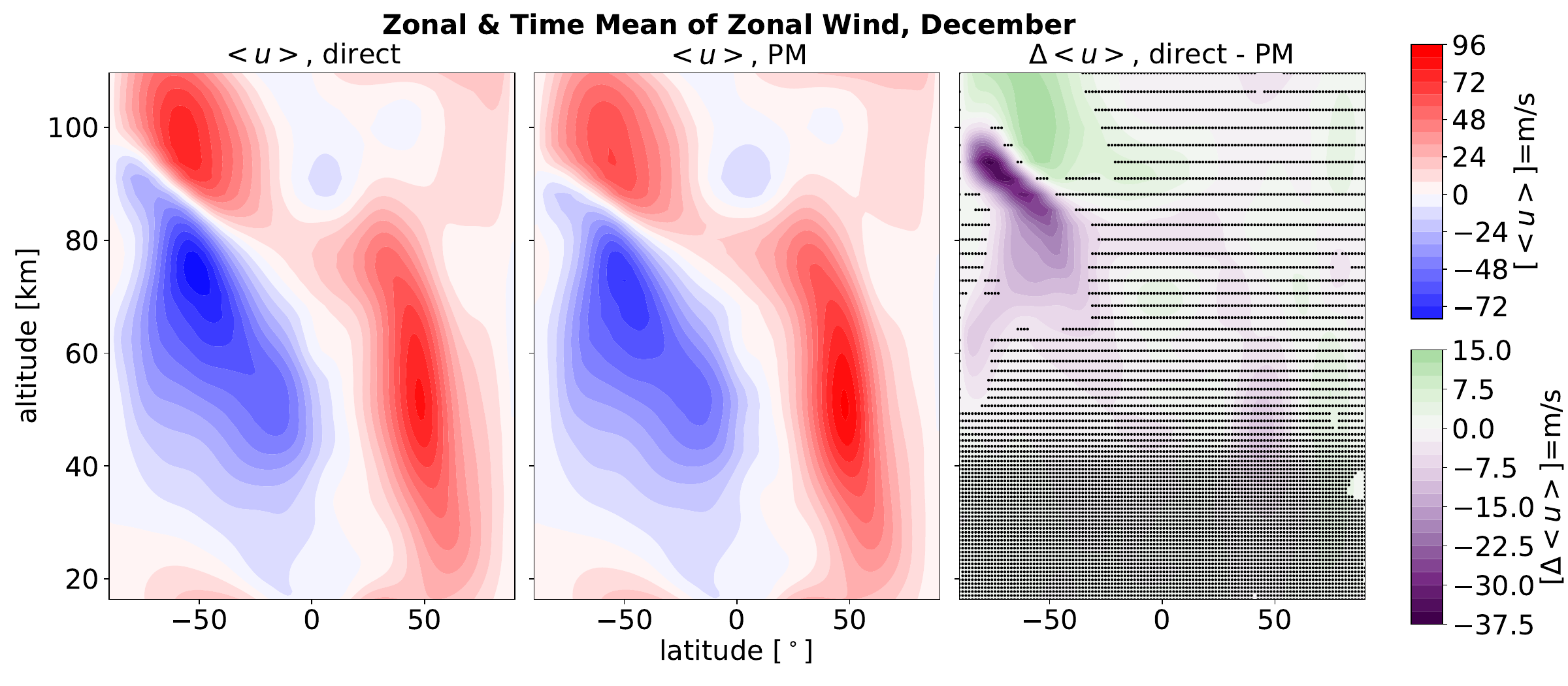}
    \caption{Climatological zonal mean zonal wind for December composites averaged over 1991-1995 and 2010-2013 (mean diurnal cycle). The left panel uses the direct approach, the middle panel uses the pseudomomentum approach, and the right panel shows the difference, direct - PM, between both approaches, with stippling indicating statistical insignificance.}
        \label{fig:pmom_compare_zonal_wind}
\end{figure}
\begin{figure}[h]
    \centering
    \includegraphics[width=\textwidth]{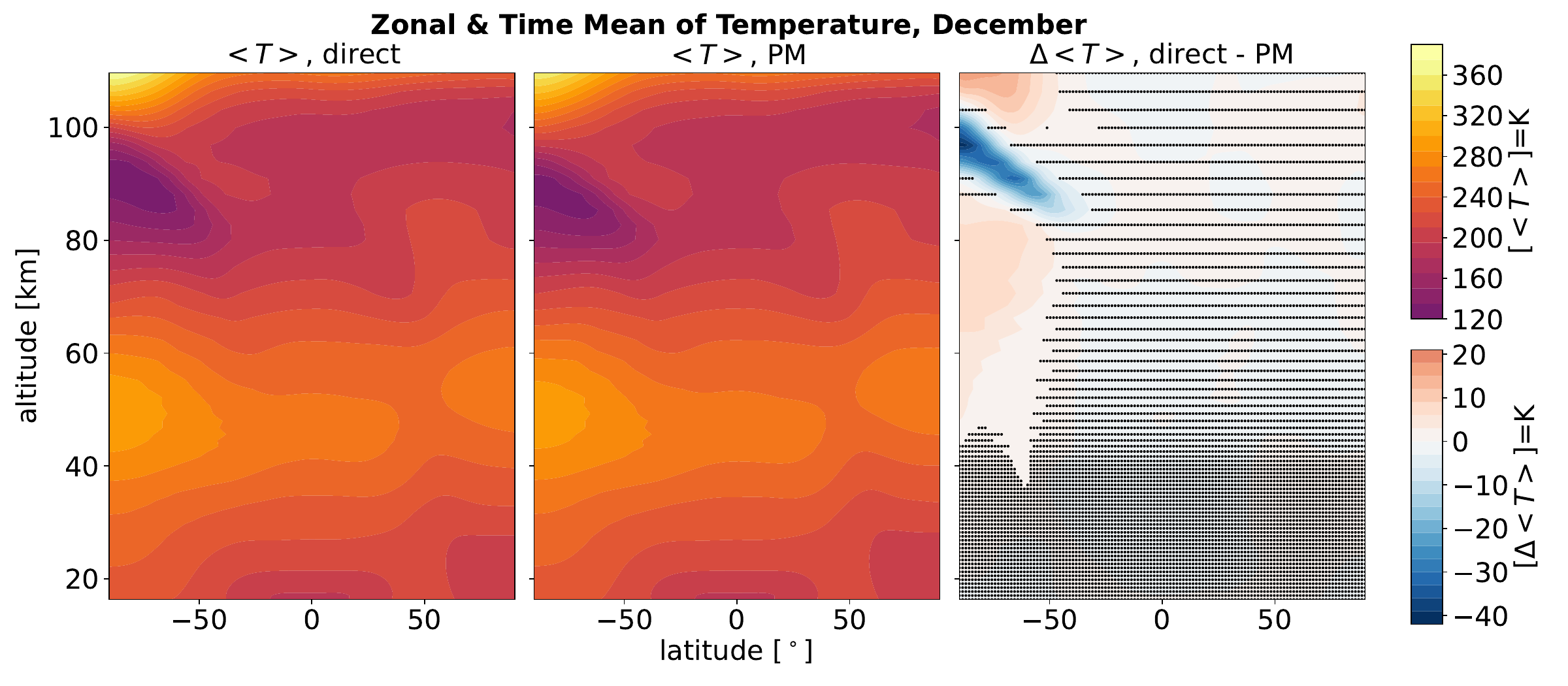}
    \caption{Climatological zonal mean temperature for December composites averaged over 1991-1995 and 2010-2013 (mean diurnal cycle). The left panel uses the direct approach, the middle panel uses the pseudomomentum approach, and the right panel shows the difference, direct - PM, between both approaches, with stippling indicating statistical insignificance.}
        \label{fig:pmom_compare_temperature}
\end{figure}

\subsection{Pseudomomentum: Direct vs. Classical Approach}
\label{subsec: Pseudomomentum: Direct vs. Classical Approach}
Today's climate models are still parameterizing GWs using the pseudomomentum approach, which simplifies the GW impact using Eliassen-Plam fluxes and further neglects thermal impact. These assumptions are only valid if the flow is in geostrophic and hydrostatic balance. Solar tides are not balanced by nature, hence one expects a significant impact on solar tides and their interaction with gravity waves when using the direct approach over PM.

For the evaluation of the effect of the direct approach as compared to PM, both in 3D, we compare two ensembles of 9 members each, initialized in November 1991-1995 and 2010-2013. Figures \ref{fig:pmom_compare_zonal_wind} and \ref{fig:pmom_compare_temperature} show results from both the pseudomomentum (left panel) and direct approach (middle panel), along with the difference direct - PM (right panel), for the December mean zonal mean zonal wind (Fig.\ref{fig:pmom_compare_zonal_wind}) and temperature (Fig.\ref{fig:pmom_compare_temperature}). Both approaches yield the wind reversal in the MLT, together with a cold summer mesopause. In the pseudomomentum approach, the mesopause is about $127.6\ \mathrm{K}$ cold and located at about $88\ \mathrm{km}$ altitude in the Southern Hemisphere at $80^{\circ}$ latitude. In the direct approach, the mesopause is shifted upward toward approximately $91\ \mathrm{km}$ altitude while being about $2.6\ \mathrm{K}$ colder than PM. Although the basic structures do not differ very much, it is interesting to see a significant effect even in winds and temperatures at high latitudes, where balance can be assumed. Both the change in amplitude and the shift of the mesopause lead to statistically significant differences in temperature and winds, whereas the shift is related to the impact of the residual-mean circulation which is ageostrophic by nature. The radial component of the residual-mean is significantly stronger in the direct approach due to gravity wave impact (not shown), shifting the mesopause upward.

\begin{figure}[t]
    \centering
    \includegraphics[width=\textwidth]{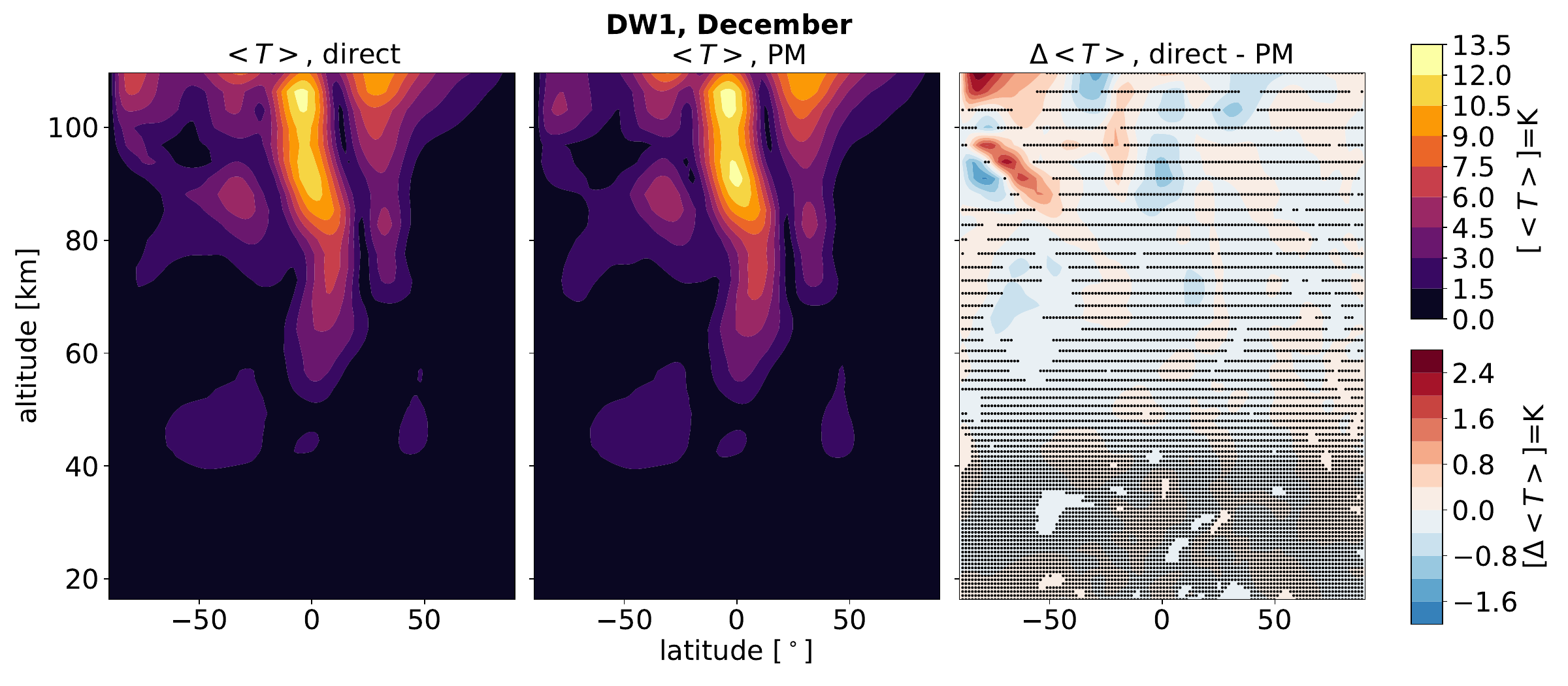}
    \caption{Migrating diurnal solar tide of the temperature for December composites averaged over 1991-1995 and 2010-2013 (mean diurnal cycle). The left panel uses the direct approach, the middle panel uses the pseudomomentum approach, and the right panel shows the difference, direct - PM, between both approaches, with stippling indicating statistical insignificance.}
    \label{fig:pmom_compare_zonal_wind_DW1}
\end{figure}

Differences in zonal mean wind and zonal mean temperature might communicate an indirect effect on STs. Figure \ref{fig:pmom_compare_zonal_wind_DW1} shows the migrating diurnal ST, i.e., DW1, in the temperature of both the direct approach (left panel) and the pseudomomentum approach (middle panel), and the difference between the two (right panel). Both approaches show similar amplitude structures, but also significant differences in the southern hemisphere's MLT region.
, where tidal signals differ up to $20\%$ of the maximum tidal signal. Other tidal components have also been analyzed. Although the migrating ST tend to differ up to $31\%$ at the lower edge of the ICON sponge at $110\ \mathrm{km}$ atltitude, non-migrating tidal signals do not differ with statistical significance (both not shown). Hence, although solar tides are not in hydrostatic and geostrophic balance, they are not as affected as assumed by using PM over the direct approach. Significant differences are located in areas with weak tidal signals and less gravity wave influence, whereas the pseudomomentum approach influences mean zonal wind and zonal mean temperature in locations with strong gravity wave influence with statistical significance. 

\begin{figure}
    \centering
    \includegraphics[width=0.95\textwidth]{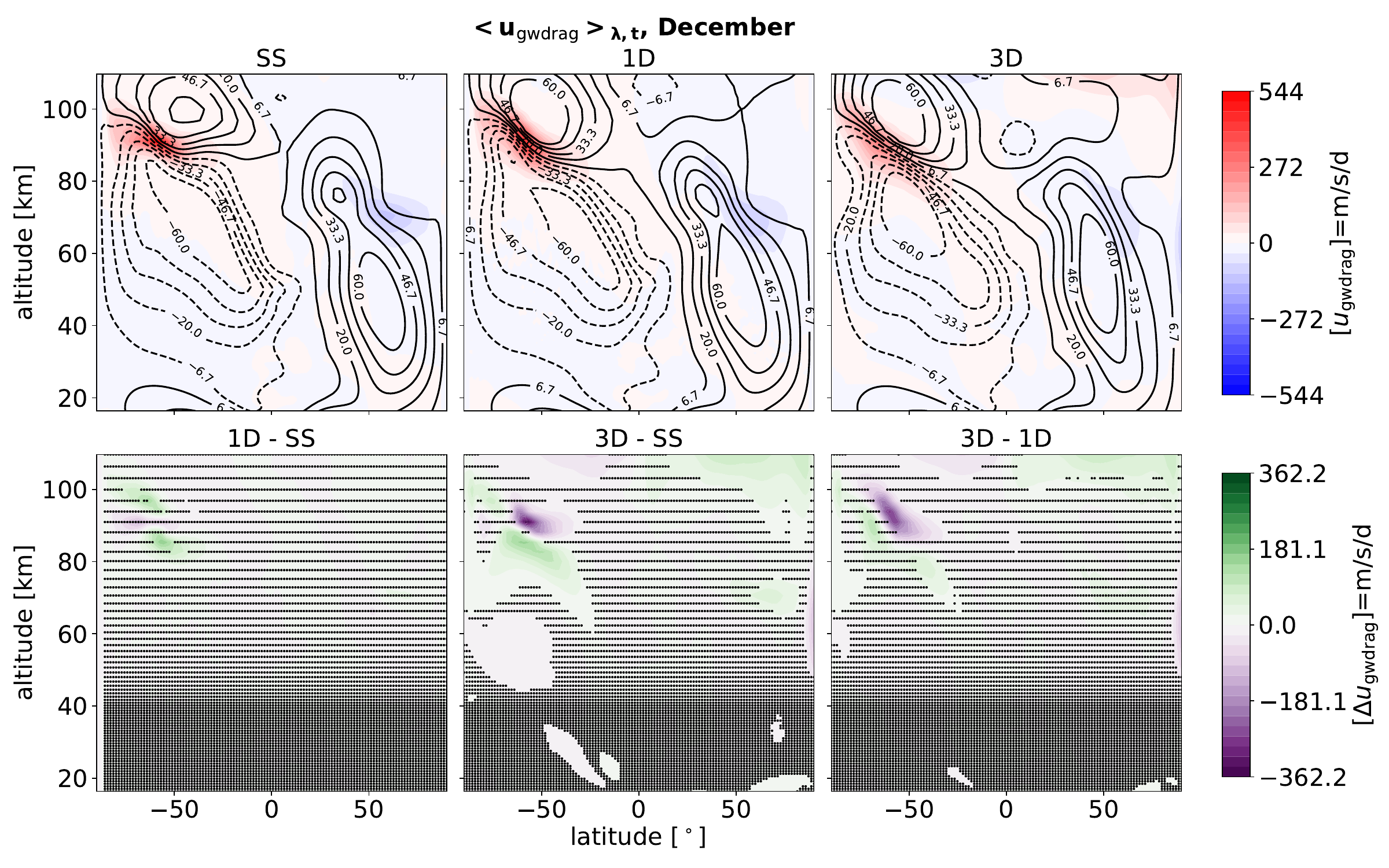}
    \caption{Climatological zonal mean zonal gravity-wave drag for December composites. The upper three plots show from left to right the results for MS-GWaM SS, 1D, and 3D, denoted by "SS", "1D", and "3D". The contour lines represent the zonal and time mean zonal wind. The lower three plots show the respective difference between the model setups, with stippling indicating statistical insignificance.
    }
    \label{fig:ICON_variable_zonal_wind_gwdrag}
\end{figure}
\begin{figure}
    \centering
    \includegraphics[width=0.95\textwidth]{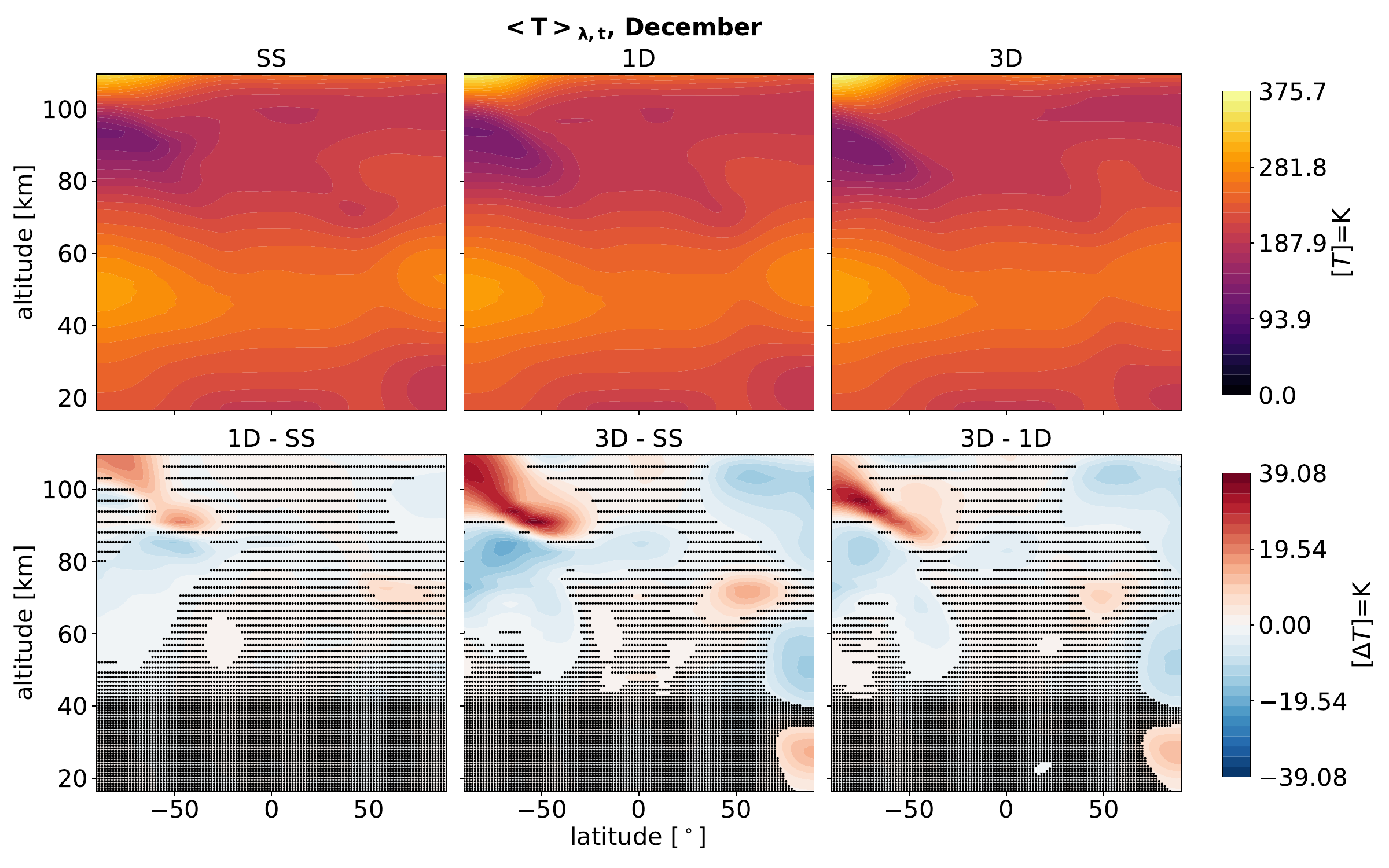}
    \caption{Climatological zonal mean temperature for December composites. The upper three plots show from left to right the results for MS-GWaM SS, 1D, and 3D, denoted by "SS", "1D", and "3D". The lower three plots show the respective difference between the model setups, with stippling indicating statistical insignificance.}
    \label{fig:ICON_variable_temperature}
\end{figure}
\begin{figure}
    \centering
    \includegraphics[width=0.95\textwidth]{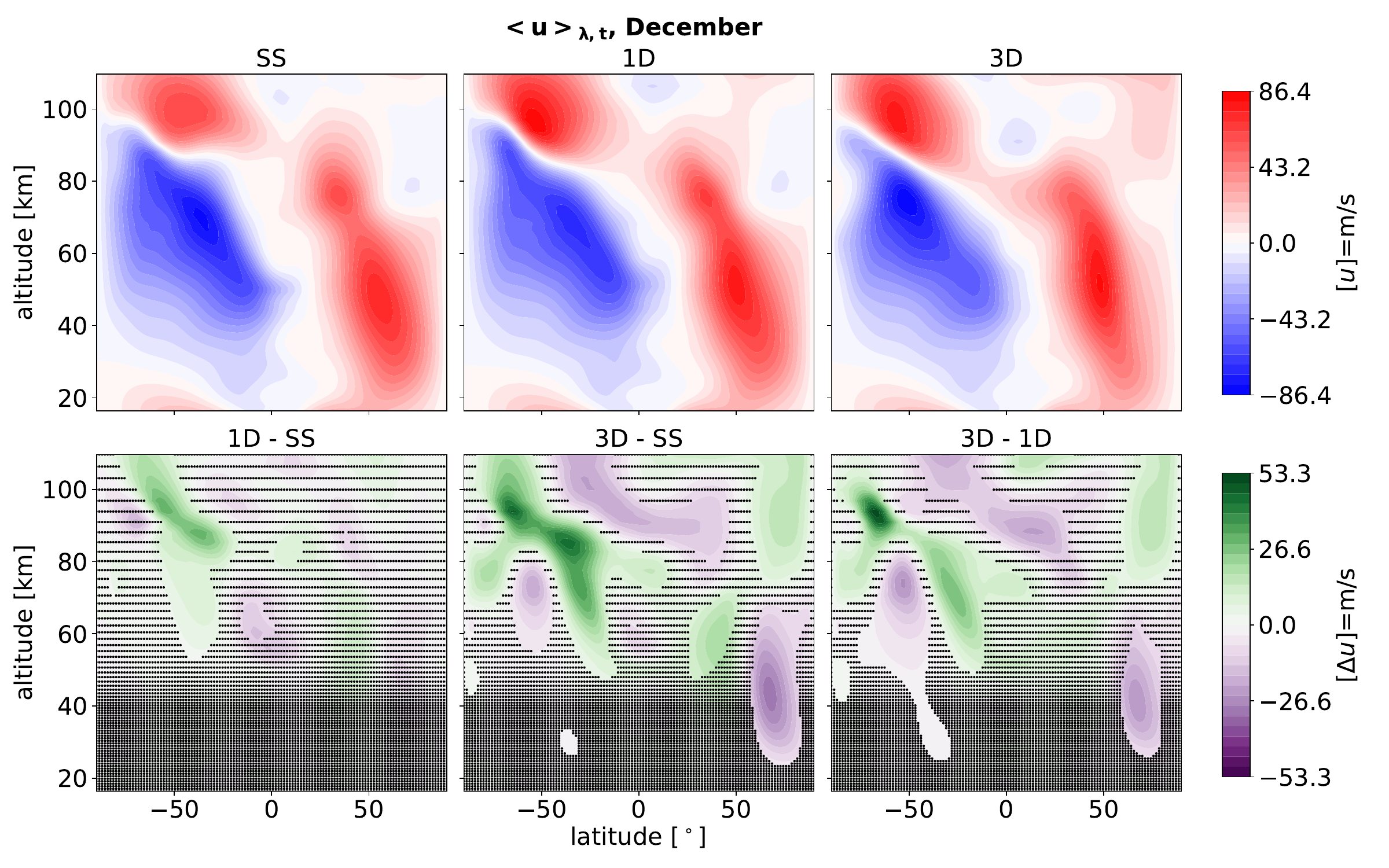}
    \caption{Climatological zonal mean zonal wind for December composites. The upper three plots show from left to right the results for MS-GWaM SS, 1D, and 3D, denoted by "SS", "1D", and "3D". The lower three plots show the difference between the model setups, with stippling indicating statistical insignificance.}
    \label{fig:ICON_variable_zonal_wind}
\end{figure}

\subsection{Effect of Transience and Horizontal Wave Propagation}
The effect of non-classical gravity-wave dynamics, i.e., the effect of transience and oblique GW propagation has been studied by \cite{ribstein2015interaction, ribstein2016interaction} with a linear tidal model. Both studies used wrong metric terms in the eikonal equations. Therefore this work can be seen as an update of those two papers, with an addinitional relaxation of the linear tidal model since solar tides are in our case part of the resolved flow.

For a comparative evaluation of the effect of oblique GW propagation and transience, we compare three ensembles using MS-GWaM in 3D, 1D and SS setup, all with the direct approach. The SS ensemble has 10 members, initialized November 1991-1995 and 2010-2014, the 1D and 3D ensemble has one member less, missing the 1994 and 2014 simulation, respectively.

\subsubsection{Zonal Mean Climatology}


Figure \ref{fig:ICON_variable_zonal_wind_gwdrag} shows the effect on the zonal-mean zonal GW drag.
While there is no statistically significant difference between the drag in steady-state and single-column model setup,
oblique GW propagation is lowering the GW drag in the summer mesopause.
This significant different GW drag (Fig. \ref{fig:ICON_variable_zonal_wind_gwdrag}) shifts the location of the summer mesopause (Fig. \ref{fig:ICON_variable_temperature}) in both altitude and latitude, which are located for MS-GWaM (SS, 1D, 3D) at about $\left(94\ \mathrm{km},\ 94\ \mathrm{km},\ 91\ \mathrm{km}\right)$ altitude and at $\left(90^{\circ},\ 87^{\circ},\ 80^{\circ}\right)$ latitude in the Southern Hemisphere, while being about $\left(120.7\ \mathrm{K},\ 122.0\ \mathrm{K},\ 125.0\ \mathrm{K}\right)$ cold. As for the comparison of GW drags of the PM vs. direct approach, the lower value in MS-GWaM 3D compared to SS and 1D leads to a warmer summer mesopause. Both transience and oblique GW propagation is shifting the summer mesopause towards the equator. The shift of the summer mesopause in altitude and latitude has an effect on the zonal wind (Fig. \ref{fig:ICON_variable_zonal_wind}). Especially oblique GW propagation, i.e., the combination of a warmer summer mesopause located at lower altitude and latitude, allows for statistical significant differences up to about $53\ \mathrm{m}/\mathrm{s}$ in the wind reversal region where GW's influence the dynamics of the atmosphere. Despite that differences in the zonal wind are related to shift of the summer mesopause in latitude and not in order of magnitude, this differences can effect the interaction between GW's and the large scale flow, and therefore also GW-ST interaction.
 This partially compensates for the effect of the using the direct momentum-fluxes. However, as interesting is that the oblique GW propagation also leads to a warming or cooling of the polar night altitudes, and to a deplacement of the polar night jets: The polar lower stratosphere is warmed, and the polar upper stratosphere and mesosphere are cooled. The polar night jet in the stratosphere is shifted polewards, while the opposite is observed in the MLT. Moreover, at low latitudes one also sees significant impacts throughout the mesosphere. Especially these might be of relevance for the tidal signal, as they direct influence tidal propagation.
\begin{figure}
    \centering
    \begin{minipage}[t]{0.48\textwidth}
        \centering
        \includegraphics[width=\textwidth]{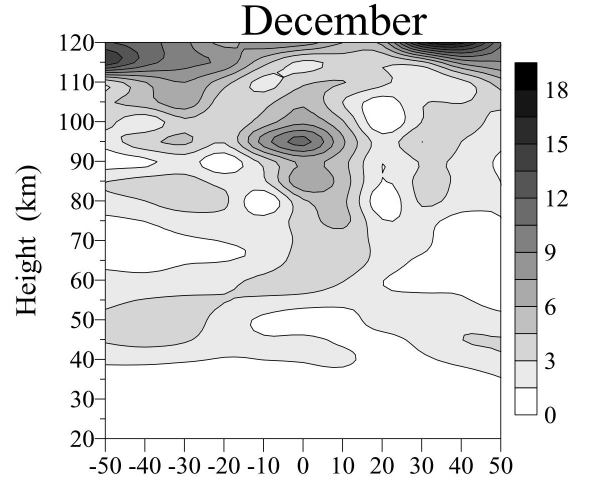}

        \textbf{(a)} Migrating diurnal solar tide, $\mathrm{DW}1$ \cite{mukhtarov2009global}
        \label{fig:Comparison_satellite_data_DW1}
    \end{minipage}
    \hfill
    \begin{minipage}[t]{0.48\textwidth}
        \centering
        \includegraphics[width=\textwidth]{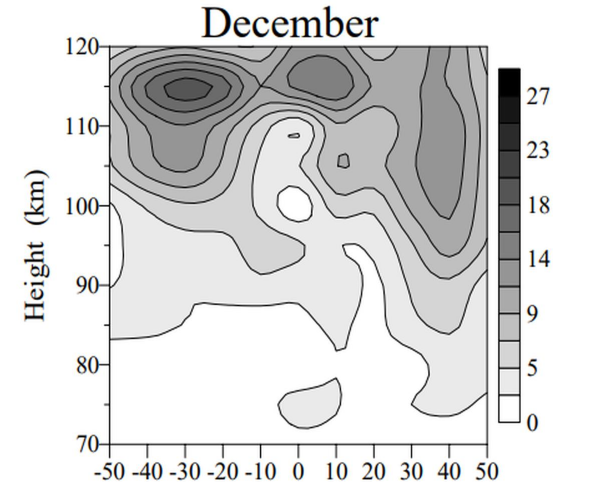}

        \textbf{(b)} Migrating semidiurnal solar tide, $\mathrm{SW}2$ \cite{pancheva2009global}
        \label{fig:Comparison_satellite_data_SW2}
    \end{minipage}

    \caption{Migrating solar tides derived from SABER/TIMED temperature measurements for December (average over 2002–2007). Panel (a) shows the diurnal solar tide $\mathrm{DW}1$ \cite{mukhtarov2009global}, panel (b) the semidiurnal solar tide $\mathrm{SW}2$ \cite{pancheva2009global}.}
    \label{fig:Comparison_satellite_data_DW1_SW2}
\end{figure}
\begin{figure}
    \centering
    \includegraphics[width=0.95\textwidth]{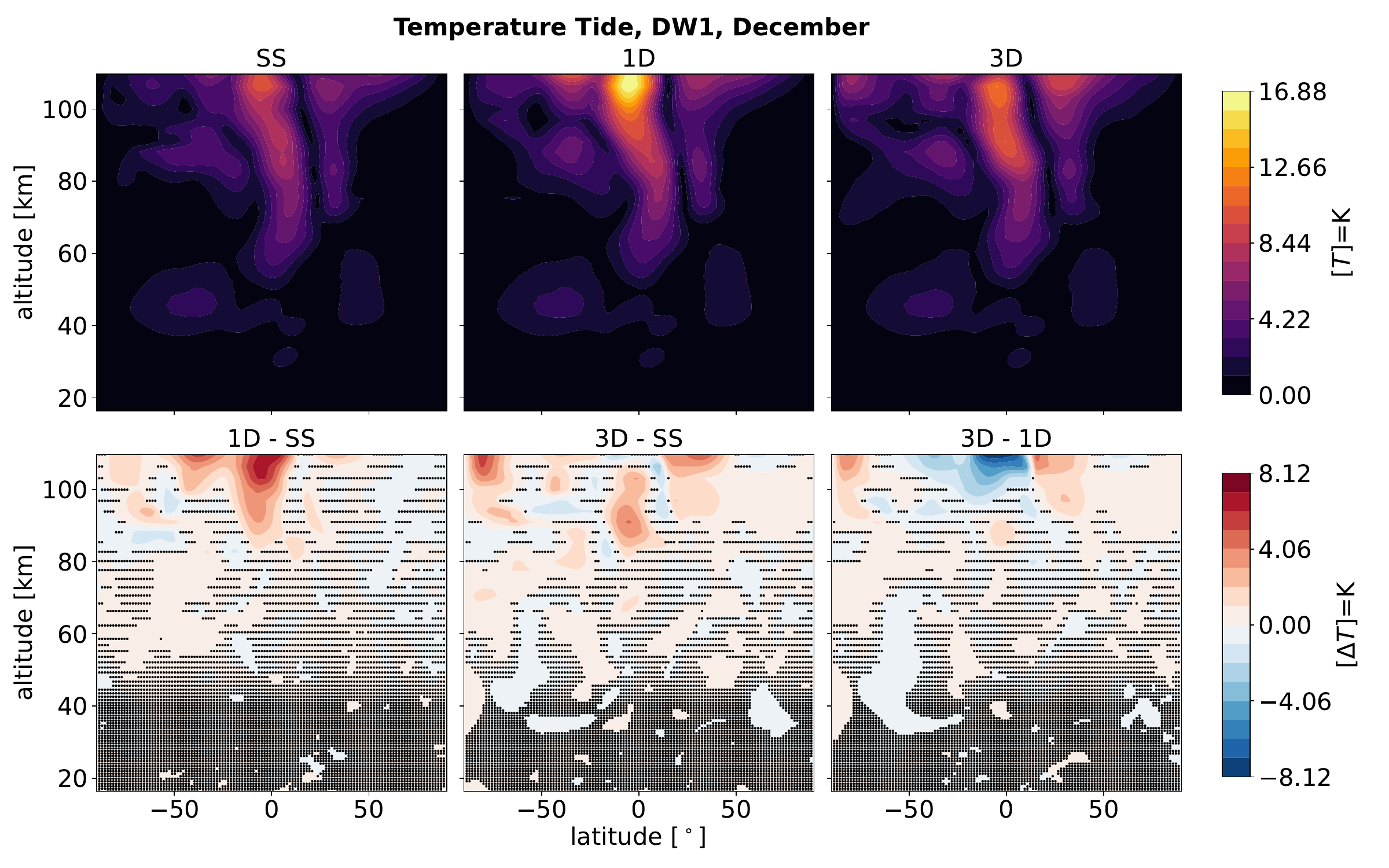}
    \caption{Migrating diurnal temperature solar tide, $\mathrm{DW}1$, for December composites. The upper three plots show from left to right the results from simulations using MS-GWaM in the modes SS, 1D, and 3D. The lower three plots show the respective difference between the model setups, with stippling indicating statistical insignificance.}
    \label{fig:Comparison_MS-GWaM_DW1}
\end{figure}

\subsubsection{Solar Tides}


The atmospheric dynamics in the MLT region are strongly influenced by solar tides and gravity waves. When investigating tidal signals for all three different MS-GWaM model setups, the GW-ST interaction changes due to the effect of transience and oblique GW propagation. Several tidal signals are compared to satellite data, to retrieve a more profound understanding of the impact of non-classical GW dynamics at the MLT region, which may later replace the classical steady state and pseudomomentum approach of state-of-the-art climate and global circulation models.

We first compare to migrating solar temperature tides retrieved from satellite data by \citeA{mukhtarov2009global} and \citeA{pancheva2009global, pancheva2013climatology}. The comparison is limited to the latitude range $\pm 50^{\circ}$ because the satellite retrievals do not give any results beyond that range.
Panel (a) of Fig. \ref{fig:Comparison_satellite_data_DW1_SW2} shows the December amplitude of the observed migrating diurnal temperature tide, published by \citeA{mukhtarov2009global}. It has a global maximum of about $12~\mathrm{K}$
at the equator, at $95\ \mathrm{km}$ altitude. In comparison to this, the maxima obtained from simulations using MS-GWaM in SS or 1D mode (left and middle panel of Fig.\ref{fig:Comparison_MS-GWaM_DW1}) are at a higher altitude of about 110km, while MS-GWaM 3D yields a maximum at about 100km altitude, closer to the observations. Moreover MS-GWaM SS leads to an underestimation of the amplitude maximum, about 12K in the observations, by about $3-4$K, while use of MS-GWaM yields an overestimation of about $4-5$K.  Only with MS-GWaM 3D, the simulated amplitude is matched well.


\begin{figure}
    \centering
    \includegraphics[width=0.95\textwidth]{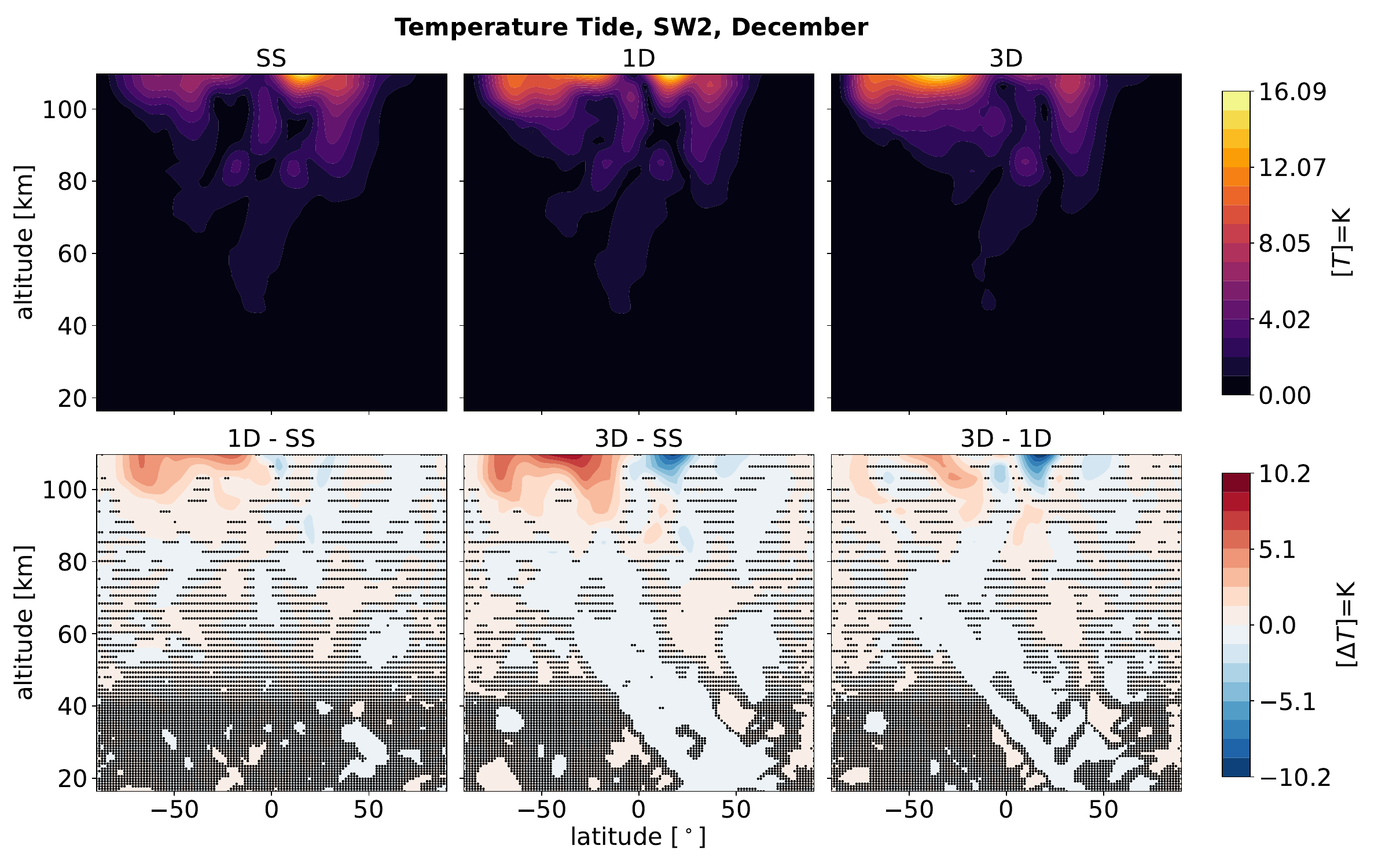}
    \caption{
    As Fig. \ref{fig:Comparison_MS-GWaM_DW1}, but now for the semidiurnal migrating tide.
    }
    \label{fig:Comparison_MS-GWaM_SW2}
\end{figure}
Next, turning to the migrating semidiurnal tide, panel (b) of Fig. \ref{fig:Comparison_satellite_data_DW1_SW2} shows its temperature amplitude retrieved from satellite data by \cite{pancheva2009global}. Only the altitude range below 110km can be compared to ICON/MS-GWaM simulation results, because this is the lower edge of the model's sponge layer. Below 110km the observed SW2 has a maximum at $30 - 40^{\circ}$S latitude with an amplitude of about 14K. Another maximum of about the same amplitude is found at $40^{\circ}$N latitude. ICON with MS-GWaM in 3D mode reproduces the two maxima approximately, but the northern-hemisphere peak is a bit too weak, as can be seen in the right upper panel of Fig. \ref{fig:Comparison_MS-GWaM_SW2}. As compared to this, MS-GWaM SS (left upper panel of Fig. \ref{fig:Comparison_MS-GWaM_SW2}) yields a global maximum in the northern hemisphere, in disagreement with the observational findings, and also MS-GWaM 1D (middle upper panel) leads to an overestimation of this feature.

Note that comparing the here-found tidal components DW2, DE2, SW4, SE2 and TW3 with retrieval from SABER/TIMED satellite data \cite{pancheva2010global,pancheva2013climatology,pancheva2024climatology} leads to similar results (not shown). The fundamental structure and amplitudes of the tidal components resulting from the simulations are generally in good agreement with the satellite observations, with MS-GWaM 3D yielding the smallest deviations.

\begin{figure}
    \centering
    \includegraphics[width=0.95\textwidth]{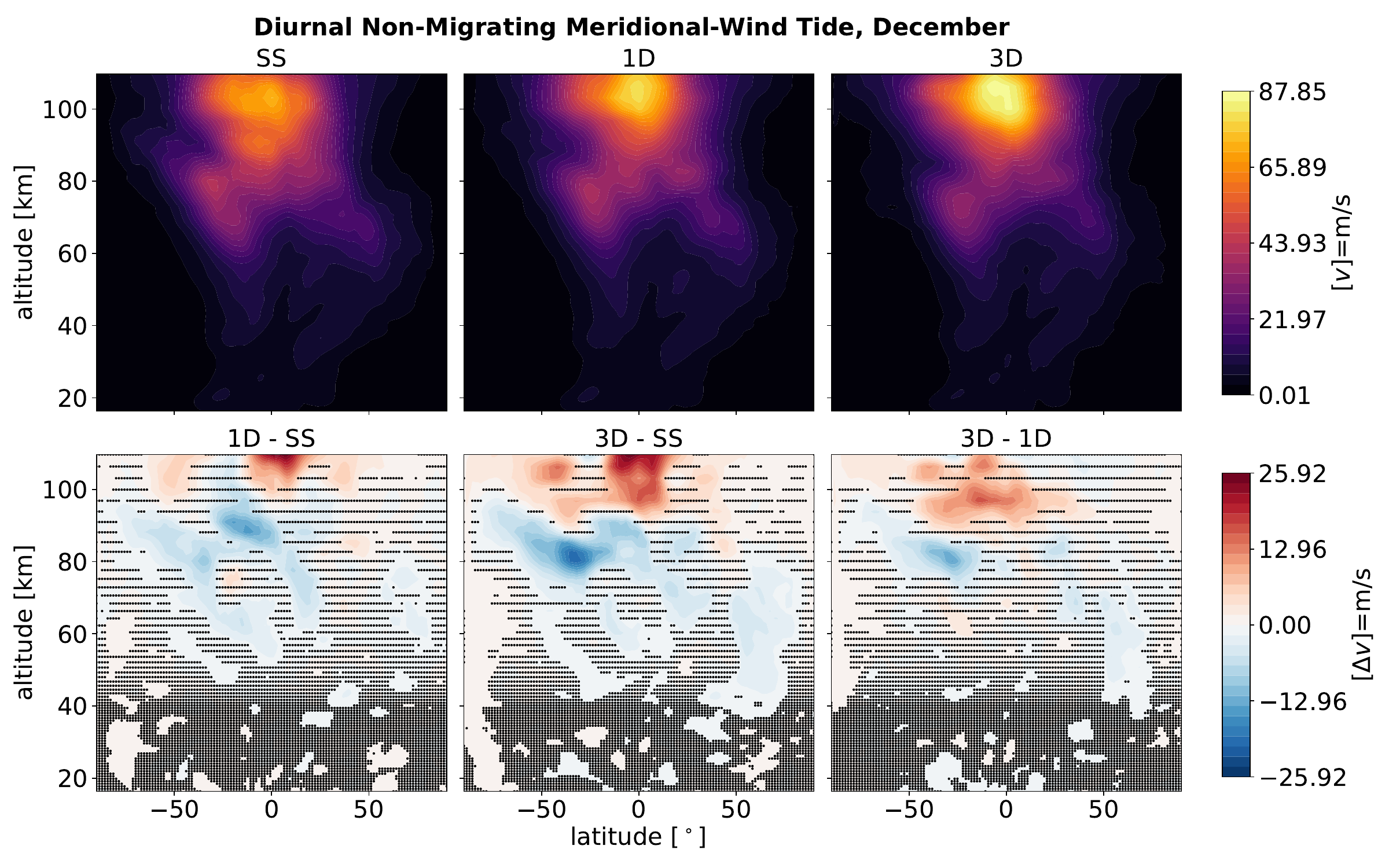}
    \caption{Non-migrating diurnal meridional-wind tide for December composites. The upper three plots show from left to right the results for MS-GWaM SS, 1D, and 3D, denoted by "SS", "1D", and "3D". The lower three plots show the respective difference between the model setups, with stippling indicating statistical insignificance.
    }
    \label{fig:Comparison_diurnal_nonmigrating_tide_meridional_wind_MSGWaM}
\end{figure}


For a comparison of the relevance of the transience of GW-mean-flow interaction on the one hand, and oblique GW propagation on the other we show in the bottom rows of Figs. \ref{fig:Comparison_MS-GWaM_DW1} and \ref{fig:Comparison_MS-GWaM_SW2} the difference between the results from using MS-GWaM in SS, 1D and 3D mode. The effect of the transience can be estimated from the difference 1D-SS (left panel), and that of oblique GW propagation from 3D-1D (right panel). The total difference between the most general and the conventional approach is indicated by the difference 3D-SS (middle panel). One sees in Fig. \ref{fig:Comparison_MS-GWaM_DW1}, that both the effect of oblique GW propagation and transience lead to significant differences up to about $\pm 8\ \mathrm{K}$ in maximum amplitude of the DW1 tidal signal. Both effects combined cancel each other out, resulting in a global maximum matching the observations using the MS-GWaM 3D mode. Also the semidiurnal migrating temperature tide is affected by both processes in the same amount. In comparison to the diurnal migrating tide, both effects of oblique GW propagation and transience rather add up in a constructive manner (see Fig. \ref{fig:Comparison_MS-GWaM_SW2}) than cancelled out (see Fig. \ref{fig:Comparison_MS-GWaM_DW1}). The same similar relevance of both effects is also seen in the migrating zonal and meridional wind tides (not shown).

\begin{figure}
    \centering
    \includegraphics[width=0.95\textwidth]{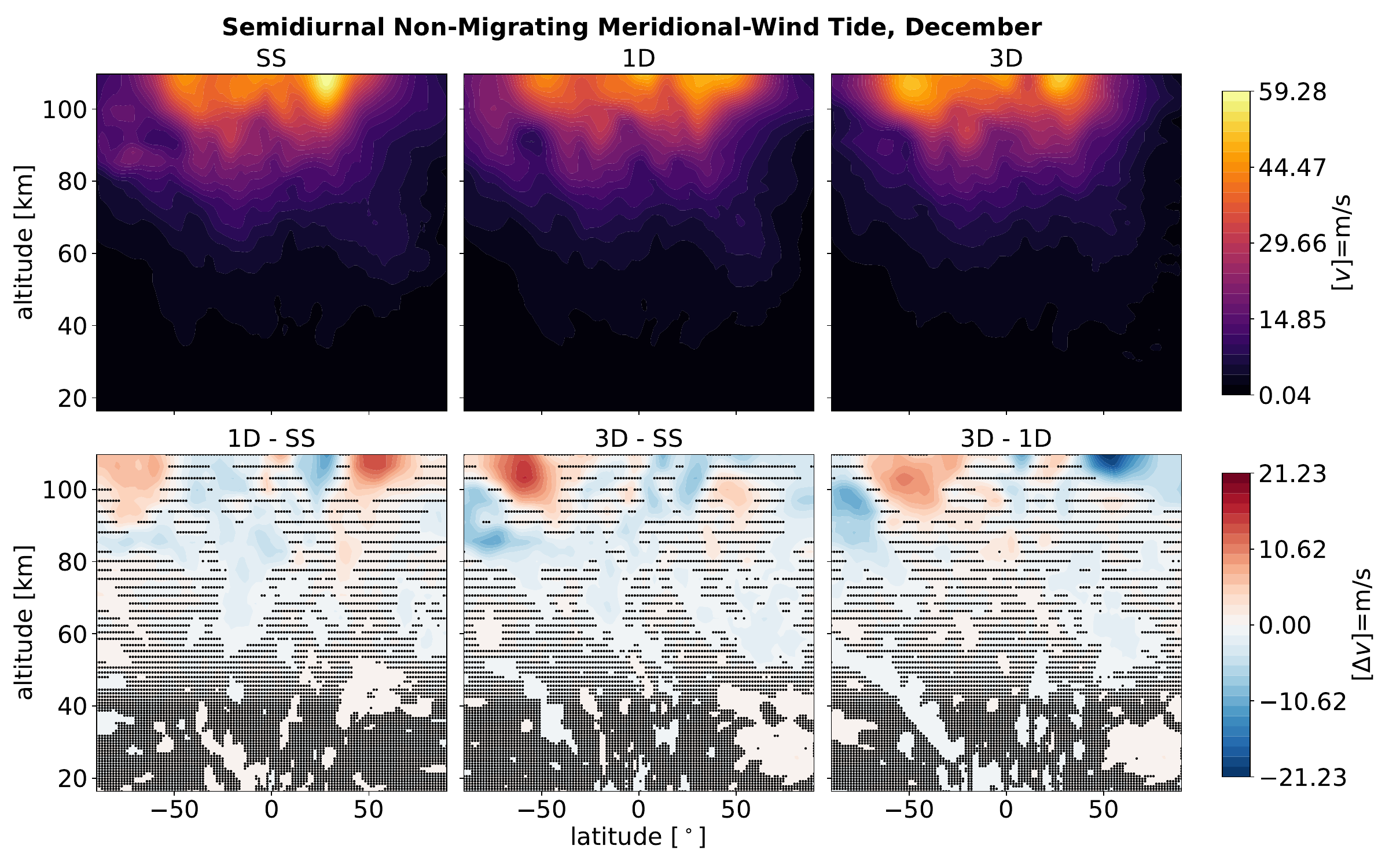}
    \caption{
    As Fig. \ref{fig:Comparison_diurnal_nonmigrating_tide_meridional_wind_MSGWaM}, but now for the semidiurnal tide.
    }
    \label{fig:Comparison_semidiurnal_nonmigrating_tide_meridional_wind_MSGWaM}
\end{figure}
\begin{figure}
    \centering
    \includegraphics[width=0.95\textwidth]{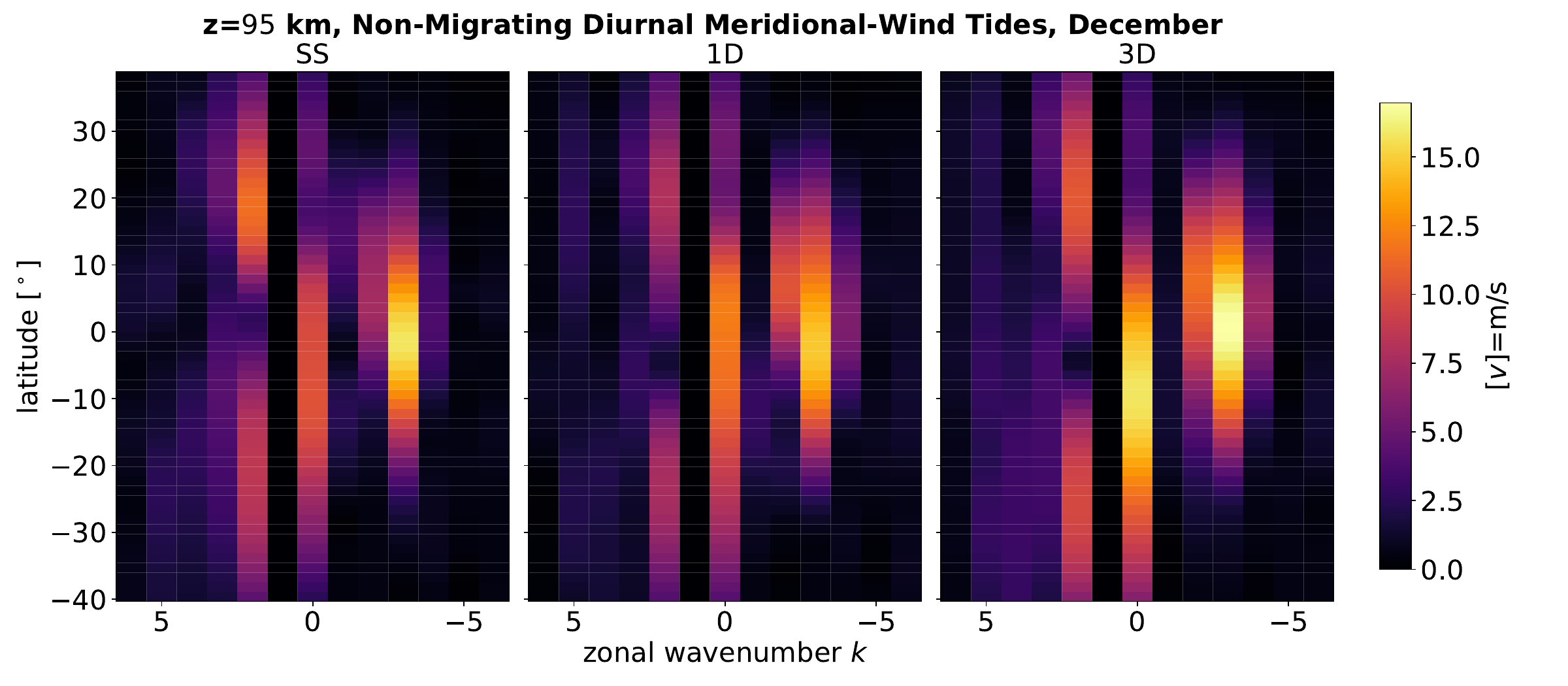}
    \caption{Power spectrum of the non-migrating diurnal meridional wind solar tides for December composites at $95\ \mathrm{km}$ altitude. The plots show from left to right the results for MS-GWaM SS, 1D, and 3D.}
    \label{fig:Comparison_diurnal_meridional_wind_nonmig_powerspectrum_MSGWaM}
\end{figure}
We finally also compare the effects on the nonmigrating tides, but for solar tides of the meridional wind instead of the temperature. Figure \ref{fig:Comparison_diurnal_nonmigrating_tide_meridional_wind_MSGWaM} depicts the results for the total (the word "total" is dropped from here on) nonmigrating diurnal meridional-wind tide (the sum of all nonmigrating components), while Fig. \ref{fig:Comparison_semidiurnal_nonmigrating_tide_meridional_wind_MSGWaM} gives the corresponding comparison for the nonmigrating semidiurnal tide.  In the case of the diurnal nonmigrating tide (Fig. \ref{fig:Comparison_diurnal_nonmigrating_tide_meridional_wind_MSGWaM}), transience (lower left panel) is most important to intensify the amplitude maximum over the equator with a maximum amplitude of about $25\ \mathrm{m}/\mathrm{s}$, but oblique GW propagation (lower right panel) also contributes to this end, but with a smaller impact indicated by an amplitude maximum of only about $15\ \mathrm{m}/\mathrm{s}$. In the case of the semidiurnal nonmigrating tide (Fig. \ref{fig:Comparison_semidiurnal_nonmigrating_tide_meridional_wind_MSGWaM}), transience (lower left panel) shifts the northern-hemisphere maximum polewards, indicated by a maximum amplitude of about $13\ \mathrm{m}/\mathrm{s}$ in the Northern Hemisphere, and oblique GW propagation (lower right panel)  reduces it, indicated by a minimum amplitude of about $21\ \mathrm{m}/\mathrm{s}$ in the Northern Hemisphere. Oblique GW propagation also establishes a second maximum in the southern hemisphere, indicated in the middle panel of Fig.(\ref{fig:Comparison_semidiurnal_nonmigrating_tide_meridional_wind_MSGWaM}) by the amplitude maximum of about $16\ \mathrm{m}/\mathrm{s}$ in the Southern Hemisphere.
\begin{figure}
    \centering
    \includegraphics[width=0.95\textwidth]{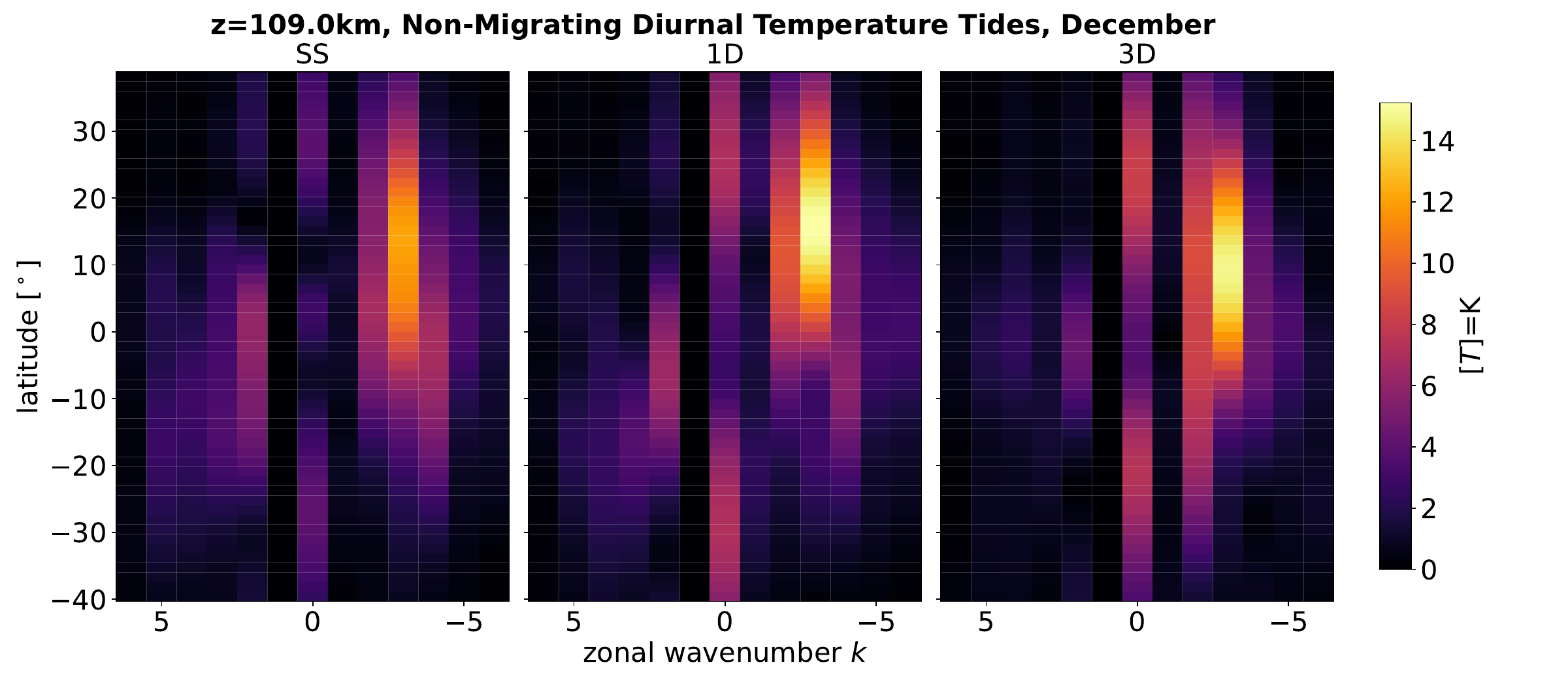}
    \caption{Power spectrum of the non-migrating diurnal temperature solar tides for December composites at $109\ \mathrm{km}$ altitude. The plots show from left to right the results for MS-GWaM SS, 1D, and 3D.}
    \label{fig:Comparison_semidiurnal_temperature_nonmig_powerspectrum_MSGWaM}
\end{figure}
Figures \ref{fig:Comparison_diurnal_meridional_wind_nonmig_powerspectrum_MSGWaM} and     \ref{fig:Comparison_semidiurnal_temperature_nonmig_powerspectrum_MSGWaM} show the spectral decomposition of the diurnal nonmigrating signal for meridional wind at 95 km altitude and temperature at 109 km altitude, respectively. All simulations reproduce the well-known dominance of the DW2, DS0, and DE3 components. Both transience and oblique GW propagation intensify the standing diurnal component DS0 and the eastward propagating component DE3. However, DW2 is weakened by transience, while it is intensified by oblique GW propagation. In all MS-GWaM modes, the DE3 tidal signal is overestimated, reaching a maximum amplitude for the respective temperature tide of about $18\ \mathrm{K}$ for MS-GWaM SS and 3D, and of about $16\ \mathrm{K}$ for MS-GWaM 1D at $95\ \mathrm{km}$ altitude.
Recent studies of \citeA<e.g.,>{li2015variability, pancheva2024climatology} on the $\mathrm{DE}3$ temperature solar tide, show that it can exceed 10K in December at altitudes of 105-110km, though it often displays lower amplitudes of about $6-8\ \mathrm{K}$ during this month. While ICON/MS-GWaM successfully captures the fundamental structure of the DE3 temperature tide, its amplitude can reach approximately 16K, thus being overestimated.

\section{Summary and Conclusions}
\label{sec: Summary and Conclusions}

Investigations of the interaction between GWs and STs have so far been limited to either linear tidal models or to conventional GW parameterizations coupled to global circulation models. Our study is the first to address this issue using a generalized GW parameterization in a global circulation model, using the Multi-Scale Gravity-Wave Model MS-GWaM, that has been implemented in the community weather and climate code ICON. As compared to conventional GW parameterizations, MS-GWaM allows the inclusion of non-classical aspects of the interaction between GWs and mean flows. These are, (i) a GW forcing of the mean flow that needs not assume the latter to be in geostrophic and hydrostatic balance, (ii) GW and mean-flow transience that does neither assume GWs propagating at infinite vertical group velocity nor mean flows that are steady during the GW propagation, and (iii) lateral GW propagation. Our investigation allows estimating the relevance of these non-classical aspects. Because gravity waves are influenced by propagation through the mean flow, it quantifies the impact not only on tidal amplitudes but also on the zonal mean zonal winds and temperatures. Only December results are given. Extensions to southern-hemisphere winter are left for future studies.



MS-GWaM compares the impact of the pseudomomentum vs. direct approach on tidal signals. Although statistically significant differences in the tidal signals between the pseudomomentum and direct approaches exist, we had expected much larger differences, since solar tides are inherently ageostrophic and thus do not satisfy the typical assumptions of geostrophic and hydrostatic balance ofthe pseudomomentum approach. The incorporation of non-classical GW forcing, i.e., the replacement of pseudomomentum (Eliassen-Palm) flux convergence in the momentum equation by momentum flux convergence and an elastic term, and the inclusion of thermal forcing by GW entropy-flux convergence leave a statistically significant impact in the summer MLT, by lifting and cooling the summer mesopause, and by an intensification of the diurnal temperature tide in the polar summer mesosphere.  Despite having weak differences in tidal signals, the lifting of the summer mesopause can be explained with the ageostrophic residual-mean circulation. Hence one should aim towards a relaxation of geostrophic and hydrostatic balance, used in the classical pseudomomentum approach, and use a more general forcing to describe the dynamics in the MLT region.


Considerable effects can be attributed to transience and lateral propagation. In comparison to simulation results using a conventional steady-state and single-column version of MS-GWaM, the zonal mean winds and temperature differ significantly throughout the mesosphere, at all latitudes, and also the polar night temperature and winds in the stratosphere exhibit significant shifts. These changes in the circulation and also the changed GW forcing lead to significant changes in the simulated tides as well. These are strongest in the MLT, but stratosphere and lower mesosphere are also affected. Both transience and oblique GW propagation contribute to this to about equal parts. Hence GW parameterizations seem to need both effects to obtain a more physical solution.

Our results naturally depend on the specific set-up of MS-GWaM, and also on the physics parameterizations in ICON in general, e.g., the radiation scheme and the handling of moisture and the latent heating it leads to. However, it is encouraging that various aspects of tidal structures reported from satellite-data retrievals are best reproduced, or approached, by using MS-GWaM in its most general configuration. Relaxing either oblique GW propagation or transience leads to a less physical solution which do not reproduce climatological tidal signals, retrieved from satellite-data, as good as MS-GWaM 3D, incorporating both transience and oblique GW propagation.

Nonetheless, many open questions remain. Issues caused by the choice of ray tracing in spherical coordinates are discussed by \citeA{MSGWaMAThreeDimensionalTransientGravityWaveParametrizationforAtmosphericModels}. The treatment of GW dynamics close to the poles still needs improvements. A numerical implementation fixing this pole problem is part of ongoing research. 

\section*{Open Research Section}
The version of ICON/MS-GWaM used to generate the datasets is available via \citeA{banerjee2025msgwam}. The datasets produced in this work have been deposited in a Zenodo repository and can be accessed via \citeA{kuehner2025impact}, which also includes the solar tide run scripts. By combining these run scripts with the model version from \citeA{banerjee2025msgwam}, the simulation can be reproduced in the solar tide setup used in this study.

\acknowledgments
U. A. thanks the German Research Foundation (DFG) for partial support through CRC 301 “TPChange” (Project No. 428312742 and Projects B06 “Impact of small-scale dynamics on UTLS transport and mixing,” B07 “Impact of cirrus clouds on tropopause structure,” and Z03 “Joint model development and modelling synthesis”), and U.A. and G.S.V. thank DFG for support through the CRC 181 “Energy transfers in Atmosphere and Ocean” (Project No. 274762653 and Projects W01 “Gravity-wave parameterization for the atmosphere” and S02 “Improved Parameterizations and Numerics in Climate Models”). 

%
%

\bibliography{bibliography}

%
%
%
%
%

\end{document}

%% file: newcommands.tex
\newcommand{\del}{\partial}

\newenvironment{myeqnarray*}
  {\begin{linenomath*}\begin{eqnarray*}}
  {\end{eqnarray*}\end{linenomath*}}

\newenvironment{myequation*}
  {\begin{linenomath*}\begin{eqnarray*}}
  {\end{eqnarray*}\end{linenomath*}}